\documentclass[twocolumn]{aastex631}

\usepackage{color}
\usepackage[normalem]{ulem} 

\begin{document}

\title{Investigating the Gamma-Ray Emission from Explosive Dispersal Outflows with Fermi-LAT}
\correspondingauthor{Paarmita Pandey, pandey.176@osu.edu}

\author[0009-0003-6803-2420]{Paarmita Pandey} 
\affil{Department of Astronomy, The Ohio State University, 140 W. 18th Ave., Columbus, OH 43210, USA}
\affil{Center for Cosmology and Astroparticle Physics, The Ohio State University, 191 W. Woodruff Ave., Columbus, OH 43210, USA}

\author{Stephen C. Lenker II}
\affil{Department of Astronomy, The Ohio State University, 140 W. 18th Ave., Columbus, OH 43210, USA}

\author[0000-0002-1790-3148]{Laura A.~Lopez}
\affil{Department of Astronomy, The Ohio State University, 140 W. 18th Ave., Columbus, OH 43210, USA}
\affil{Center for Cosmology and Astroparticle Physics, The Ohio State University, 191 W. Woodruff Ave., Columbus, OH 43210, USA}

\author[0000-0003-4423-0660]{Anna L.~Rosen}
\affil{Department of Astronomy, San Diego State University, San Diego, CA 92182, USA}
\affil{Computational Science Research Center, San Diego State University, San Diego, CA 92182, USA}

\author[0000-0001-9888-0971]{Tim Linden}
\affil{Stockholm University and The Oskar Klein Centre for Cosmoparticle Physics, Alba Nova, 10691 Stockholm, Sweden}

\author[0000-0003-2377-9574]{Todd A.~Thompson}
\affil{Department of Astronomy, The Ohio State University, 140 W. 18th Ave., Columbus, OH 43210, USA}
\affil{Center for Cosmology and Astroparticle Physics, The Ohio State University, 191 W. Woodruff Ave., Columbus, OH 43210, USA}
\affil{Department of Physics, Ohio State University, 191 W. Woodruff Ave, Columbus, OH 43210}

\author[0000-0003-1252-9916]{Stella S.~R.~Offner}
\affil{Department of Astronomy, University of Texas at Austin, Austin, TX, 78712, USA}

\author[0000-0002-4449-9152]{Katie~Auchettl}
\affiliation{School of Physics, The University of Melbourne, Parkville, VIC 3010, Australia}
\affiliation{Department of Astronomy and Astrophysics, University of California, Santa Cruz, CA 95064, USA}

\author[0000-0002-2951-4932]{Christopher~M.~Hirata}
\affil{Department of Astronomy, The Ohio State University, 140 W. 18th Ave., Columbus, OH 43210, USA}
\affil{Center for Cosmology and Astroparticle Physics, The Ohio State University, 191 W. Woodruff Ave., Columbus, OH 43210, USA}
\affil{Department of Physics, Ohio State University, 191 W. Woodruff Ave, Columbus, OH 43210}

\begin{abstract}
We present the first systematic study of explosive dispersal outflows (EDOs) as potential sources of high-energy emission in the Milky Way. EDOs are energetic outflows produced during dynamical interactions in young, massive star-forming regions, and their physical conditions make them promising environments for cosmic-ray acceleration. Using 16 years of $0.2-500$ GeV Fermi-LAT observations, we study the gamma-ray properties of seven EDOs. Three EDOs, DR21, G34.26$+$0.15, and G5.89$-$0.39 show spatially coincident GeV emission, while the remaining systems yield non-detections. Among the sample, DR21 stands out as the brightest candidate, with a detection significance $\geq40\sigma$. Its spectrum is well described by a power law with an exponential cutoff, and the integrated luminosity in the 0.1$-$500 GeV band is $L_\gamma \simeq 2\times10^{35}$ erg s$^{-1}$. When compared with the outflow’s estimated kinetic energy, the inferred cosmic-ray acceleration efficiency is $\leq 15\%$, consistent with values for shocks in dense molecular environments. The energetics and morphology support an association between the DR21 molecular outflow and the observed gamma rays. Our results demonstrate that EDOs span a wide range of gamma-ray luminosities and efficiencies, suggesting they may contribute to the Galactic cosmic ray budget. This motivates searches for additional EDOs and improved multiwavelength characterization of their environments.
\end{abstract}


\section{Introduction} \label{sec:intro}

Massive star-forming regions are confirmed efficient particle accelerators \citep{Aharoniannature, Padovani_2020, Padovani_2021, peron2025}. Several young massive star clusters have been detected in the GeV energy range with the Large Area Telescope found on the Fermi Gamma-ray Space Telescope (Fermi-LAT) \citep{Cygnus, Cyg2, Westerlund2, NGC36031, NGC36032, Liu22, W40, W43, Pandey2024, Peron24, Ge2024}. Although stellar winds are considered the primary source of gamma-ray emission in young ($\leq3$ Myr) star-forming regions \citep{Bykov2020}, particle acceleration may also occur in other types of sources, such as molecular outflows \citep{Pado2016, Gaches2016}.  

To date, two distinct types of molecular outflows have been identified in these environments \citep{Frank2014, Bally2016}. The first class of sources is protostellar jets from individual young stars \citep{Arce2007}. {Gamma-ray emission from two protostellar outflows has been observed to date \citep[HH 80-81;][]{Yan2022, Emma2025} and \citep[S255 NIRS 3;][]{emma2023}}. The second class consists of explosive dispersal outflows (EDOs). These are believed to result from the dynamical disruption of a young, massive, non-hierarchical stellar system, potentially triggered by the merger of massive protostars or by collisions between forming stars \citep{Zapata_orion, Zapata_2017}. 

Several key morphological and kinematic features distinguish EDOs from classical bipolar molecular outflows \citep{Zapata_2017}.  EDOs consist of straight, narrow, CO filament-like ejections with varying orientations and an almost isotropic spatial distribution, with the filament orientations converging back to a common origin, presumed to be the site of the explosive event. The radial velocity of each filament increases linearly with its projected distance from the origin in a Hubble-like flow, and the redshifted and blueshifted filaments frequently overlap in the plane of the sky. These outflows are typically associated with regions of high-mass star formation, and the flows have substantial kinetic energies of $10^{47-49}$~erg \citep{Zapata_2017}. 

{Currently, seven Galactic EDOs associated with massive star-forming regions have been confirmed through their distinctive molecular-gas kinematics: Orion  Becklin-Neugebauer/Kleinmann-Low (BN/KL) \citep{Zapata_orion, Bally_orion, Bally2020}, Sh$106-$IR \citep{Bally_SH106}, G$5.89-0.39$ \citep{Zapata_G589}, IRAS $16076-5134$ \citep{Guzm_IRAS5134}, IRAS $12326-6245$ \citep{Zapata_123}, G$34.26+0.15$ \citep{Issac2025}, and one associated with DR21 \citep{Zapata_dr21, Guzman2024ALMA}.
A study by \cite{Guzm_IRAS5134} estimates that such events occur roughly once every $\sim110$ years across the Milky Way. Remarkably, this frequency is comparable to both the Galactic core-collapse supernova rate \citep[$\sim$ one in fifty years;][]{Tammann1994} and the formation rate of massive stars \citep[also $\sim$ one in fifty years;][]{Mckee1997, Chomiuk_2011}. Taken together, these findings imply that EDOs could represent a common, short-lived phase in the evolution of massive star-forming regions.}

{In this paper, we perform a systematic study of $\gamma$-ray emission from EDOs using 16 years of Fermi-LAT data. In our analysis, we find the EDO DR21 to be the most significant detection ($\geq 40\sigma$ in the $0.2-500$ GeV range), so we will focus on this source as well as describing the full population.} 

We have organized this paper as follows: in Section~\ref{sec: analysis}, we present the sample of EDOs (Section~\ref{sec: sample}), their Fermi $\gamma$-ray analyses, including spatial and likelihood analysis (in Sections~\ref{sec: likelihoodanalysis}). We found significant gamma-ray detection from DR21, G$34.26+0.15$ and G$5.89-0.39$, so we further proceeded to study their spatial extent in Section~\ref{sec: extensionanalysis} and spectral energy distributions in Section~\ref{sec: spectralanalysis}.  In Section~\ref{sec:discussion}, we calculate the particle acceleration efficiency of the outflows assuming a hadronic $\gamma$-ray origin. We estimate the contribution of EDOs to the galactic cosmic ray (CR) budget and discuss the implications. In Section~\ref{sec:DR21_gamma-ray}, we discuss the properties of DR21, evaluate the association between the EDO and the observed gamma-ray emission, and consider alternative emission scenarios. {In Section~\ref{Sec:Conclusions}, we summarize the conclusions.}

\begin{deluxetable*}{cccccc}
\tablenum{1}
\tablecaption{List of all the EDOs in our sample.
}
\tablewidth{240pt}
\tablehead{
\colhead{Source Name} &   
\colhead{Distance\tablenotemark{a}}  &
\colhead{Age (yr)\tablenotemark{b}}  &
\colhead{Number Density\tablenotemark{c}} &
\colhead{Kinetic Energy\tablenotemark{d}} &
\colhead{Galactic Coordinates (l,b) } 
}
\startdata
G$34.26+0.15$ & $3.3 \pm 0.3$ kpc & $\leq$19000 & $10^{5}$ $\rm cm^{-3}$ & $10^{48}$ erg & (34.25, $+$0.15) \\
DR21\tablenotemark{$\dag$} & $1.5 \pm 0.08$ kpc & 10000 & $10^{4-6}$ $\rm cm^{-3}$ & $10^{48}$ erg & (81.527, $+$0.543) \\
Sh $106-$IR & $1.09\pm 0.05$ kpc & 3500 &  $ 10^4$ $\rm cm^{-3}$ & $10^{47}$ erg & (76.36, $-$0.59)     \\
IRAS $16076-5134$ & $5.0\pm 0.7$ kpc &  3500 & $-$ & $10^{48-49}$ erg & (331.28, $-$0.18)  \\
G$5.89-0.39$\tablenotemark{$\dag$} & $2.99 \pm 0.19$ kpc & 1000 & $10^5$ $\rm cm^{-3}$ & $10^{46-49}$ erg &  (5.887, $-$0.392) \\
IRAS $12326-6245$ &  $2.03\pm0.77$ kpc & 700 & $-$& $10^{48}$ erg & (301.136, $-$0.225)  \\
Orion BN/KL & $388 \pm 5$ pc & 500 & $10^{4-6}$ $\rm cm^{-3}$ & $10^{47}$ erg & (208.51, $-$20.27)  \\
\enddata
\tablenotetext{a}{Distance values taken from \cite{Kucher1994, Rygl2012,  Zucker_sh106, Baug2020_5134, Sato_g05, Duronea_123, Kounkel_orion}, respectively.}
\tablenotetext{b}{Age refers to the kinematic or dynamical age of the EDOs, which estimates how long ago the outflow event occurred, based on its observed size and velocity. Taken from \cite{Issac2025, Zapata_dr21, Bally_SH106, Guzm_IRAS5134, Zapata_G589, Zapata_123, Bally_orion}, respectively.}
\tablenotetext{c}{Number density taken from \cite{Issac2025, Jakob_2006, Schenider2007_s106,  Stark_2007, Peng_2012} respectively.}
\tablenotetext{d}{Kinetic energy taken from \cite{Issac2025, Zapata_dr21, Bally_SH106, Guzm_IRAS5134, Zapata_G589, Zapata_123, Bally_orion}, respectively.}
\tablenotetext{$\dag$}{Coincident Fermi sources are 4FGL J2038.4+4212, 4FGL J1800.2$-$2403c, respectively.}
\label{tab:table_sample}
\end{deluxetable*}

\section{Sample, Data Analysis and Results \label{sec: analysis}}

\subsection{Sample \label{sec: sample}}

{Table~\ref{tab:table_sample} lists the sample of EDOs that we study. They span a range of distances ($0.4-5$ kpc), dynamical ages ($\sim 500-19000$ yr), and outflow kinetic energies ($10^{46}-10^{49}$~erg). These estimates were taken from recent high-resolution studies of molecular gas dynamics \citep[e.g.;][]{Zapata_orion, Zapata_dr21, Guzman2024ALMA}. Ages correspond to the dynamical expansion times inferred from the velocity field of the molecular gas, while the kinetic energies are estimates of the total mechanical energy of the outflow.}

Two members of the sample, DR21 and G$5.89–0.39$, coincide spatially with catalogued Fermi-LAT sources (4FGL J$2038.4+4212$ and 4FGL J$1800.2–2403$c, respectively), making them the most promising targets for our analysis. The sources exhibit a variety of physical environments ranging from compact, young systems like Orion BN/KL to more evolved regions such as DR21. 

\subsection{Data Selection \label{sec: spatialanalysis}}

We use data taken by the Large Area Telescope (LAT) on board the Fermi Gamma-Ray Space Telescope. The LAT uses multiple layers of conversion foil for background rejection and a calorimeter to measure the energies of incoming $\gamma$-ray photons. It can detect $\gamma$-rays in the range of 0.1 $-$ 500 GeV, with a field of view of roughly 2.4 steradians, a resolution of $\approx 8^{'}$ at energies exceeding 2 GeV, and an effective area of 9,500 cm$^2$ at normal incidence \citep{Atwood2009}. 

{We analyze 16~years of Pass~8 \texttt{SOURCE}-class $\gamma$-ray data 
(from 2008~August~4 to 2025~September~9; MET 239846401$-$780245937) in the 
0.2$-$500~GeV energy range. For each EDO listed in Table~\ref{tab:table_sample}, we used a $20^{\circ} \times 20^{\circ} $ square region centered on the source. To reduce contamination from $\gamma$-rays produced by CR interactions in the Earth’s atmosphere, we excluded photons with zenith angles greater than $90^{\circ}$. Only good time intervals (GTIs) were retained by applying the standard quality filters \texttt{DATA\_QUAL > 0} and \texttt{LAT\_CONFIG == 1}. The analysis was performed using the \texttt{FermiPy} Python package \citep[v1.4.0;][]{Wood_fermipy} and ScienceTools version 2.2.0. Events were restricted to the \texttt{SOURCE} event class (\texttt{evclass = 128} and \texttt{evtype = 3}). The instrument response functions \texttt{P8R3\_SOURCE\_V3} were applied to analyze the SOURCE events. We employed a binned maximum-likelihood analysis within each $20^{\circ}\times 20^{\circ}$ image, using eight energy bins per decade and an angular pixel size of $0.1^{\circ}$.  The background model included all sources from the most recent \texttt{4FGL-DR4} catalog \citep{Abdollahi_2022, ballet2024} as well as the standard Galactic diffuse emission model (\texttt{gll\_iem\_v07.fits}) and isotropic background (\texttt{iso\_P8R3\_SOURCE\_V3\_v1.txt}).}

\subsection{Likelihood Analysis \label{sec: likelihoodanalysis}}

We performed a binned maximum likelihood analysis of the EDOs using \texttt{FermiPy} to measure the $\gamma$-ray emission. This method estimates the best-fit parameters for a given model of $\gamma$-ray sources and their spectra by maximizing the joint likelihood across spatial and spectral bins. The likelihood function $\mathcal{L}$ represents the probability of obtaining the observed data given a specific model. We defined the test statistic (TS) as $\text{TS} = -2 \ln(\mathcal{L}_0 / \mathcal{L}_1)$, where $\mathcal{L}_0$ and $\mathcal{L}_1$ are the likelihoods of models without and with an additional source at the center of our image, respectively. {We allowed the normalizations of the Galactic and isotropic diffuse backgrounds to vary, along with all the normalizations of all sources within 5$^\circ$ of the target. Because many sources lie on the Galactic plane, freeing the normalization beyond 5$^\circ$ caused the likelihood fit to fail, so we fixed it at 5$^\circ$, which produced a stable, convergent fit. Sources located beyond this radius, and those with TS $\leq$ 25, were fixed to their 4FGL-DR4 values. For the EDOs DR21 and G$5.89-0.39$, we assumed the 4FGL sources were their $\gamma-$ray counterparts for the analysis. For the other 5 EDOs (G$34.26+0.15$, Sh $106-$IR, IRAS $16076-5134$, IRAS $12326-6245$, and Orion BN/KL), we added a point source power-law (PL) model at their coordinates for the analysis. We did not get any significant gamma-ray detection from Sh $106-$IR, IRAS $16076-5134$, IRAS $12326-6245$, and Orion BN/KL, so we report their $\gamma-$ray luminosities from $2\sigma$ flux upper limits.
We got significant values of TS = 10.11 and 32.13 for G$34.26+0.15$ and G$5.89-0.39$ respectively. For DR21, we found TS = 1199.62. For these three sources, we report the estimated $\gamma-$ray luminosities.

\begin{figure}[t]
\begin{center}
\includegraphics[width=\columnwidth]{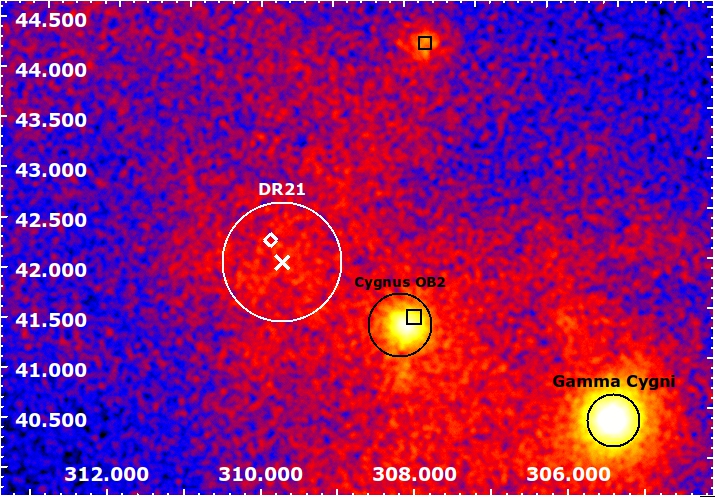}
\caption{Fermi-LAT $2-500$~GeV counts map of the $6^{\circ}\times4^{\circ}$ region surrounding DR21. The white X and the surrounding white circle show the position and extent of 4FGL~J$2038.4+4212$ (see Section~\ref{sec: extensionanalysis}), the proposed $\gamma$-ray counterpart, and the white diamond marks the DR21 EDO. Black squares mark PSR~J2032$+$4127 and PSR~2030$+$4415, and the black circles indicate the locations of the Cygnus OB2 association and Gamma Cygni.}
\label{fig:DR21_countsmap}
\end{center}
\end{figure}

\subsubsection{Likelihood Analysis for DR21}\label{sec:likelihood_DR21}

In the Fermi data, the 4FGL J$2038.4+4212$ source is offset 0.154$^\circ$ ($\approx 9'$) from DR21, and thus it is a possible candidate for $\gamma$-ray emission from DR21. According to the Fermi 4FGL-DR4 catalog, this source has a TS value of 1199.62 when using a log-parabola spectrum. There are two pulsars within 3$^\circ$: PSR J2032$+$4127 and PSR 2030$+$4415 are located 1.54$^\circ$ and 2.44$^\circ$ away, respectively \citep{Manchester2005}. Since the Fermi-LAT PSF is $\leq 1^{\circ}$ at 1~GeV, we can distinguish the emission due to the pulsars from that of DR21. 

In Figure~\ref{fig:DR21_countsmap}, we show the $2-500$ GeV counts map over a $6^{\circ} \times 4^{\circ}$ region, highlighting the positions of 4FGL J2038.4$+$4212 and DR21. {We used this energy range because Fermi-LAT’s spatial resolution above 2 GeV is approximately $18\arcmin$, which is sufficient to resolve distinct emission structures.} Based on a spatial correlation and likelihood analysis, we find DR21 is the most probable dominant $\gamma$-ray emitter associated with 4FGL J2038.4$+$4212. {We used the \texttt{gta.lightcurve()} method to generate the light curve for this source using a bin size of 182 days ($\approx$ half year) to check its variability and possible association with a variable gamma-ray source. Figure~\ref{fig:lc_dr21} shows the light curve of 4FGL J$2038.4+4212$. No significant variability was apparent over the time period of 16 years. }

\begin{figure}[t]
    \begin{center} 
    \includegraphics[width=\columnwidth]{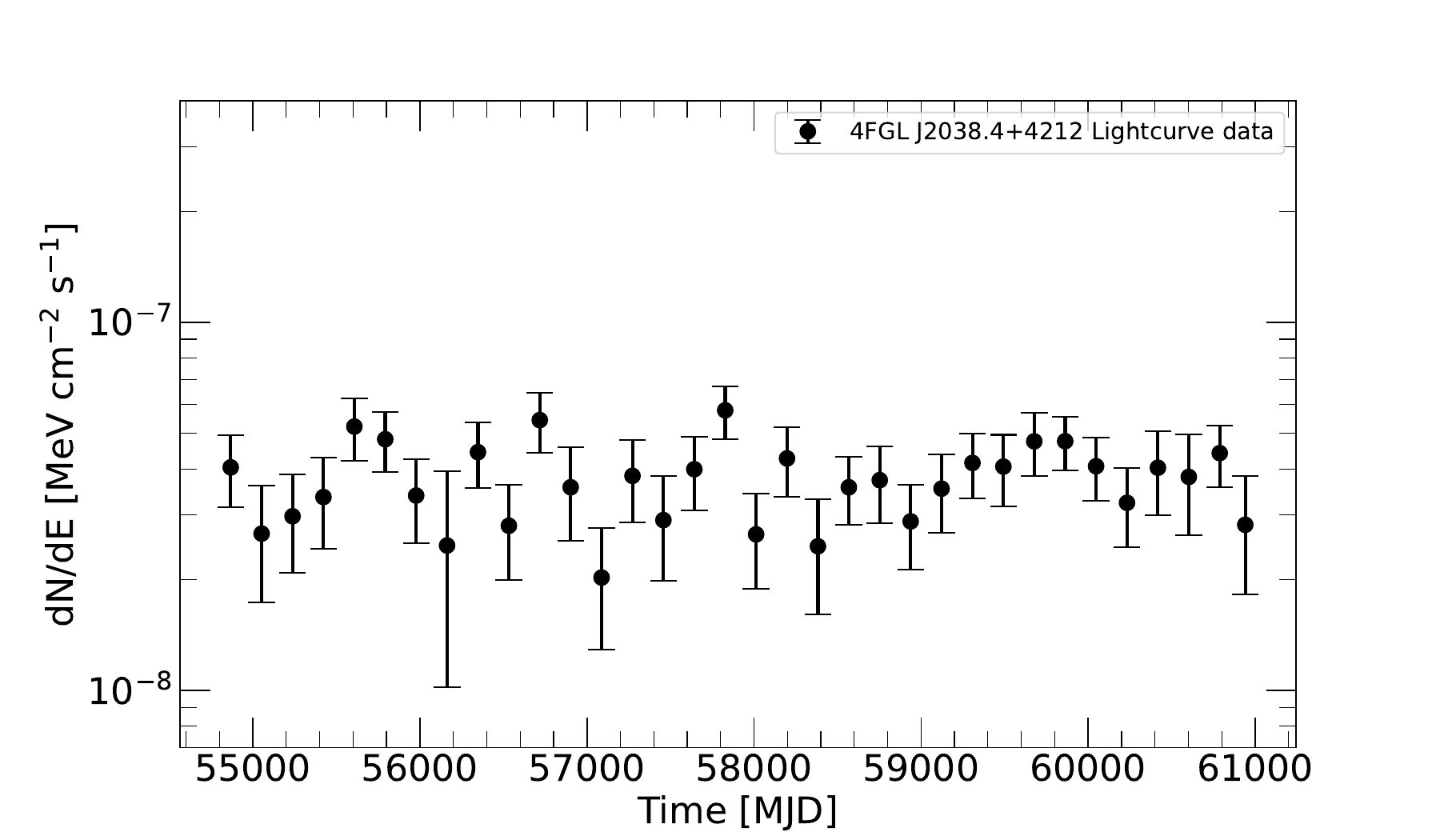}
        \caption{Fermi $\gamma$-ray light curve of 4FGL~J2038.4+4212. The light curve spans 16 years of data with a bin size of 182 days ($\sim$ half year). No significant variability was detected in flux during this time period. }  
       \label{fig:lc_dr21}
    \end{center}  
\end{figure}

To investigate the association of the Fermi source with DR21 and to study the inherent distribution of accelerated particles, we modeled the $\gamma$-ray spectrum of 4FGL J2038.4$+$4212. Other than the 4FGL catalog-adopted Log-Parabola (LP) model, we tested a power-law (PL) and power-law with an exponential cutoff (PLEC) model. The PL model is defined as
\begin{equation}   
    \frac{dN(E)}{dE} = N_0 \left(\frac{E}{E_{\rm p}}\right)^{\Gamma},
\end{equation}   
where $\Gamma$ is the spectral index, $N_0$ is the normalization (with units of ph~cm$^{-2}$~$\rm{s}^{-1}$~MeV$^{-1}$), and $E_{\rm p}$ is the pivot energy, chosen as the energy at which the error on differential flux is minimal. The LP model is defined as
\begin{equation}
   \frac{dN(E)}{dE} = N_0 \left( \frac{E}{E_p}\right)^{-(\alpha + \beta {\rm log} \frac{E}{E_p})},
\end{equation}
where $N_0$ is again the normalization and $\alpha$ and $\beta$ are the spectral index and curvature parameters, respectively. The PLEC model is defined as
\begin{equation}   
    \frac{dN(E)}{dE} = N_0 \left(\frac{E}{E_{\rm p}}\right)^{\Gamma_1} \rm exp\left( - \left(\frac{E}{E_{\rm c}}\right)^{\Gamma_2}\right),
\end{equation} 
where there are two power-law indices $\Gamma_1$ and $\Gamma_2$, $E_{\rm p}$ is the pivot energy, and $E_{\rm c}$ is the cutoff energy. 

\begin{deluxetable*}{lcccc}
\tablenum{2}
\tablecaption{List of different models used for spectral and spatial analysis and their corresponding Log-likelihood values for the $0.2-500$ GeV band. Models use the 4FGL-DR4 catalog.
}
\tablewidth{0pt}
\tablehead{\colhead{Source Position} &  \colhead{Spectral Model} &
\colhead{Source Type} & \colhead{$\Delta$AIC\tablenotemark{a}} & \colhead{TS}
}
\startdata
4FGL~J2038.4$+$4212 & Log-Parabola & Point source & 99.04 & 1199.62  \\
4FGL~J2038.4$+$4212 & Power-Law & Point source & 111.05 & 986.65  \\
Physical Coordinates of DR21\tablenotemark{b} & Power-Law & Point source & 245.93 & 889.21 \\
4FGL~J2038.4$+$4212 & PLSuperExpCutoff & Point source & 0 & 1075.33  \\
\hline \\
4FGL~J2038.4$+$4212 & Log-Parabola & Point source &  478.2 & 1199.62  \\
4FGL~J2038.4$+$4212 & PLSuperExpCutoff & Radial Disk &  47.39 & 1448.84\\
4FGL~J2038.4$+$4212 & PLSuperExpCutoff & Radial Gaussian & 0 & 1696.11\\
\enddata
\tablenotetext{a}{  $\Delta$AIC=
$\rm AIC_m - AIC_{min}$ is the difference in AIC between each model $m$ and the one that minimizes the AIC ($\Delta$AIC=0 for the best
available model).}
\tablenotetext{b}{R.A., DEC= (309.757, 42.327)}
\label{tab:TS_values_dr21}
\end{deluxetable*}

{To compare the relative quality of different spatial or spectral models, we use the Akaike Information Criterion \citep[AIC;][]{AKC1974}.  The AIC is defined as
\begin{equation}
\mathrm{AIC} = 2k - 2\ln L ,
\end{equation}
where $k$ is the number of free parameters in the model and $\ln L$ is the maximum log-likelihood of the fit. For two competing models, the one with the smaller AIC value is statistically preferred.}  

{Table~\ref{tab:TS_values_dr21} gives the different spatial and spectral models used, fitted extensions, $\Delta$AIC values, and TS values of the fits for each analysis. Here, $\Delta$AIC= $\rm AIC_m - AIC_{min}$ is the difference in AIC between each model $m$ and the one that minimizes the AIC ($\Delta$AIC=0 for the best available model). The PLEC model is the best description compared to the others. Therefore, we continued our analysis further while maintaining the source as a PLEC at the position of 4FGL~J2038.4$+$4212.}

\subsection{Extension Analysis \label{sec: extensionanalysis}}

{We ran an extension analysis with the {\fontfamily{cmtt}\selectfont {GTAnalysis.extension}} method to investigate if the EDOs DR21, G$34.26+0.15$ and 
G$5.89-0.39$ are best characterized as extended sources. We used the {\it Radial Disk} and {\it Radial Gaussian}, where the width of the extended source is defined by $\sigma$ and the radius is defined by $R$. The isotropic background and galactic background were free parameters for our analysis. We did not get any significant detection of extension for EDO G$34.26+0.15$. For EDO G$5.89-0.39$, we find the extension for a radial Gaussian template is $0.69^\circ \pm 0.07^\circ$ ($\rm TS_{ext} = 26.58$), but this may be the result of possible contamination from the Galactic background since it is close to the Galactic plane (see coordinates in Table~\ref{tab:table_sample}). Therefore, we model this source as a point source.} For DR21, we find that the best-fitting model is the Radial Gaussian. The best central position for this extended source is RA $\approx 309.707^\circ$ and Dec $\approx +42.101^\circ$, which is $\approx 7.5'$ from the 4FGL-DR4 location. The Gaussian width is $\sigma = 0.65^\circ\pm0.03^\circ$, and the TS value of extension compared to the point source model is $\text{TS}\approx 498$ (see Table~\ref{tab:TS_values_dr21} for reference). The new overall TS of the source is $\approx 1696 ~(\geq 40\sigma)$ in the $0.2-500$ GeV band when modeled as a PLEC Radial Gaussian extended source.

To examine the spatial distribution further, we produced a TS map of the $2 - 500$ GeV emission, as this energy range of Fermi-LAT has the best spatial resolution (the spatial resolution of Fermi-LAT is $\sim8'$ at 2 GeV). We used the {\tt fermipy gta.tsmap} method to produce this TS map, and we adopted the best-fit model output by {\fontfamily{cmtt}\selectfont {GTAnalysis.extension}} method, removing the source associated with DR21. Figure~\ref{fig:tsmap_3color}(a) gives the resulting TS map of DR21. The highest TS value of $\approx$~70 is in the central region and corresponds to an $\approx$~8$\sigma$ detection in the 2$-$500 GeV band. 

\begin{figure}[t]
\centering
\begin{minipage}{0.45\textwidth}
  \includegraphics[width=\textwidth]{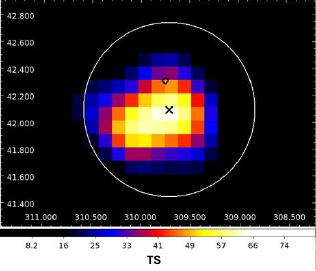}
  \centering
  (a)
\end{minipage}
\hfill
\begin{minipage}{0.45\textwidth}
  \includegraphics[width=\textwidth]{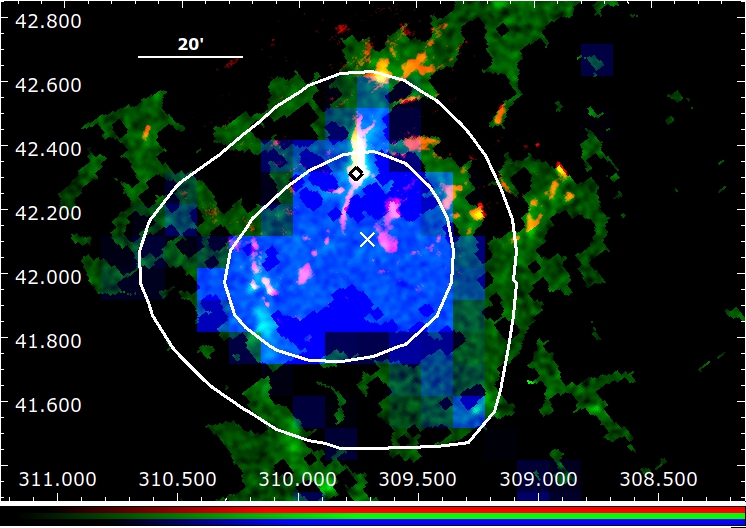}
  \centering
  (b)
\end{minipage}
\caption{
(a) TS map of the 2$-$500~GeV band data using a pixel size of $0.1^{\circ}\times0.1^{\circ}$. The maximum TS value of $\approx70$ in the central pixels is spatially coincident with the star cluster and corresponds to $\approx8\sigma$ detection in the 2$-$500 GeV band. The black diamond corresponds to the physical coordinates of the DR21 outflow, and the black X shows the new best-fit position at RA $\approx 309.707^\circ$ and Dec $\approx +42.101^\circ$. The white circle is the total size of the extended $\gamma$-ray emission region of radius $\approx 0.65^\circ$. (b) Multiwavelength three-color image of the DR21 region, with 1.1-mm radio continuum data in red from the Bolocam Galactic Plane Survey (BGPS) \citep{Ginsburg2013}, the integrated CN map from the Nobeyama 45m Cygnus-X CO survey \citep{Yamagishi2018Nobeyama} in green, and the Fermi-LAT TS map in blue. The radio emission, CN maps, and the $\gamma$-ray emission are all spatially coincident. The white contours show the $5\sigma$ (TS~$=25$; inner contour) and $3\sigma$ (TS~$=9$; outer contour) $\gamma$-ray detections.}
\label{fig:tsmap_3color}
\end{figure}

Next, we want to place this $\gamma$-ray emission in the broader context of the DR21 environment and to assess its spatial correlation with dense molecular gas. Figure~\ref{fig:tsmap_3color}(b) compares the Fermi $\gamma$-ray TS map with the 1.1 mm continuum emission from the Bolocam Galactic Plane Survey (BGPS) \citep{Ginsburg2013} and the integrated CN line emission from the Nobeyama 45m Cygnus-X CO survey \citep{Yamagishi2018Nobeyama}. The 1.1 mm emission traces thermal radiation from cold dust, associated with dense star-forming cores and embedded protostars \citep{Enoch_2006}, while CN is a high-density tracer sensitive to gas with number densities of $\sim 10^5$ cm$^{-3}$ \citep{SchiandLeroy2024}. We find that the peak of the $\gamma$-ray emission is spatially coincident with both the DR21 outflow and surrounding dense gas structures traced by CN and dust continuum emission. This spatial correspondence suggests that the particle acceleration responsible for the $\gamma$-rays may be occurring within or close to dense molecular material, potentially providing the target material for hadronic interactions. These provide important implications for the origin of the high-energy emission, which we discuss in Section~\ref{sec:DR21_gamma-ray}.

\subsection{Spectral Analysis \label{sec: spectralanalysis}}

{For DR21, we used {\tt fermipy gta.sed()} to compute the SED of the extended source using eight flux bins.  Figure~\ref{fig:sed_dr21} shows our integrated $\gamma$-ray spectrum, along with the best-fit PLEC model and its error bars. The green histogram gives the TS value of each flux bin. The two energy bins above 40 GeV have TS $<10$, so we show a 2$\sigma$ upper limit instead. The best fit PLEC model parameters are $N_0 = (4.13 \pm 0.19)\times 10^{-10}$, $\Gamma_1 = -2.01 \pm 0.04$, $\Gamma_2 = 1.50$, $E_{\rm c} = 5.31 \pm 0.73~\mathrm{GeV}$, and  $E_{\rm p} = 200~\mathrm{MeV}$ ($\Gamma_2$ and $E_{\rm p}$ were fixed parameters). The $\gamma$-ray flux over the 0.1$-$500 GeV range is $\Phi^{> 100 \rm MeV}_{\gamma} = (8.38\pm0.28) \times 10^{-8}$ ph cm$^{-2}$ s$^{-1}$. Assuming a 1.5 kpc distance, the total $\gamma$-ray luminosity is $L_\gamma \simeq (1.71\pm0.15)\times10^{35}$ erg s$^{-1}$}. 

\begin{figure}[t]
    \begin{center} 
    \includegraphics[width=\columnwidth]{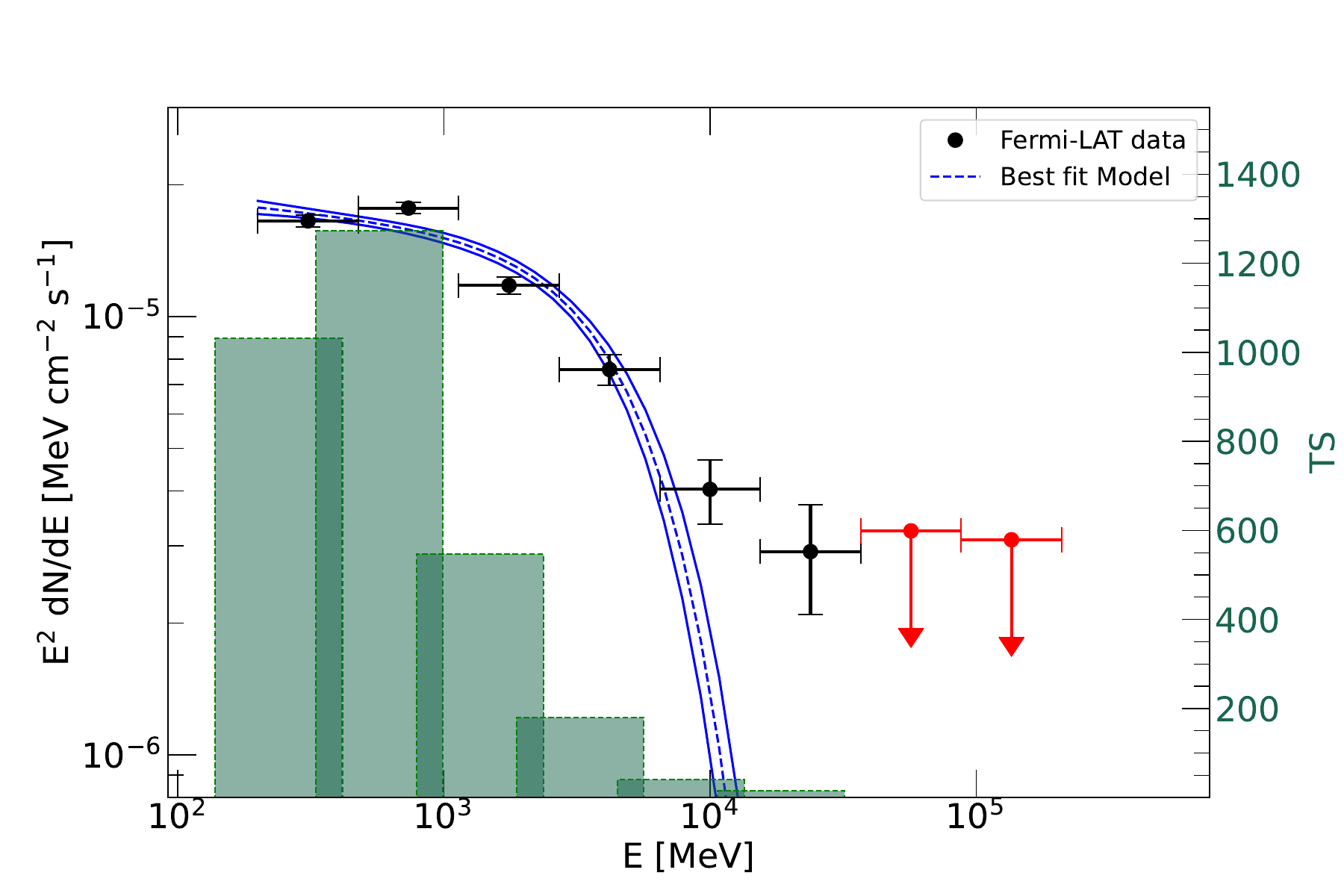}
        \caption{Fermi $\gamma$-ray SED of DR21. For each data point, the error bar reflects the statistical uncertainty caused by the effective area, and the red data points are $2\sigma$ upper-limits. The green histogram is the TS value of each flux bin. The blue dashed line shows the PLEC model with best-fit values of $\Gamma_1 = -2.01 \pm 0.04$ and $E_{\rm c} = 5.31 \pm 0.73~\mathrm{GeV}$. The solid blue lines are the 1-$\sigma$ error bars of the spectral fit. The $\gamma$-ray flux in the 0.1$-$500 GeV range is $\Phi^{> 100 \rm MeV}_{\gamma} = (5.88\pm2.89) \times 10^{-8}$ ph cm$^{-2}$ s$^{-1}$. Assuming a 1.5 kpc distance, we find the total $\gamma$-ray luminosity to be $L_\gamma \simeq (1.71\pm0.15)\times10^{35}$ erg s$^{-1}$.}  
       \label{fig:sed_dr21}
    \end{center}  
\end{figure}

\begin{figure*}[t]
    \centering    
    \includegraphics[width=0.45\textwidth]{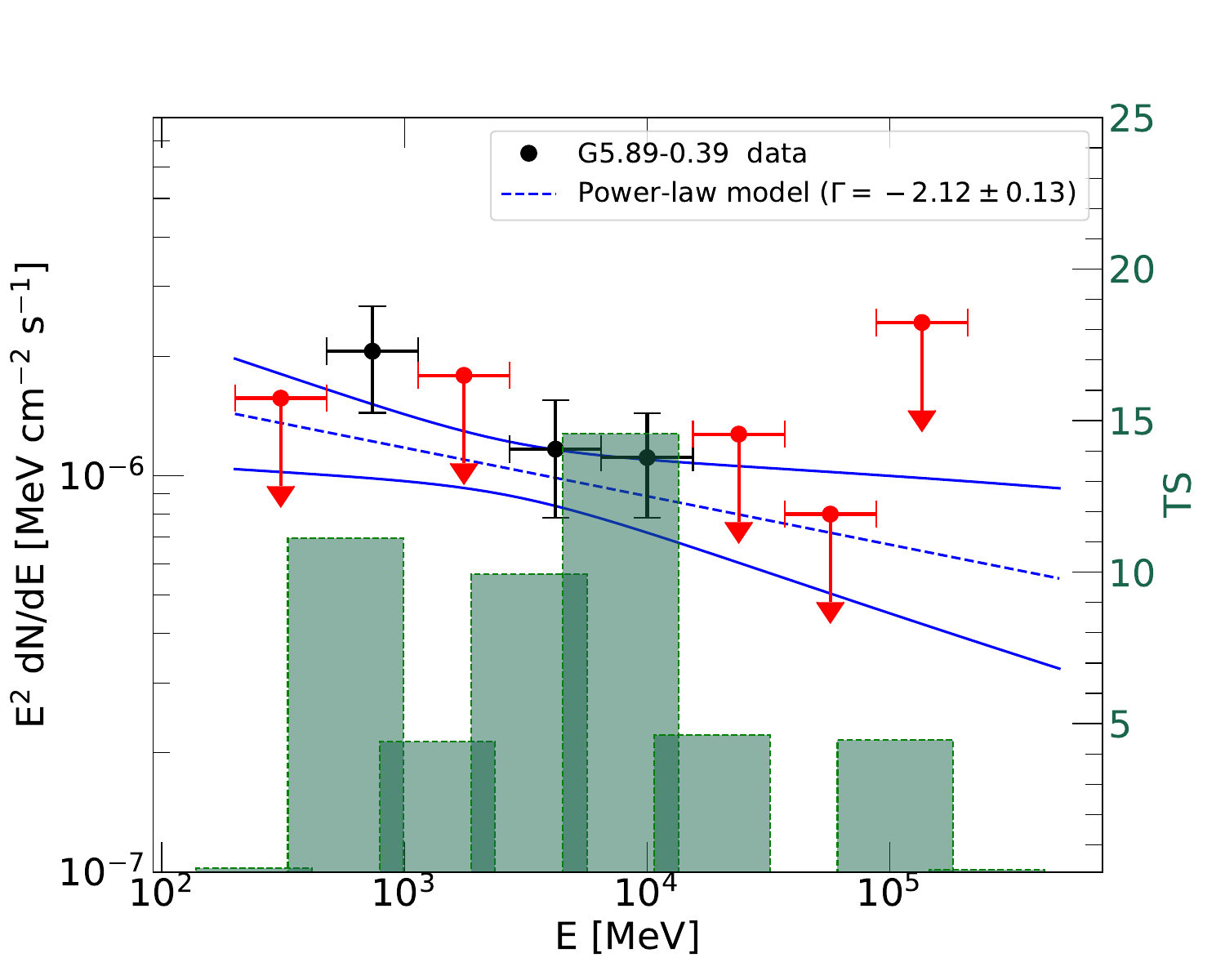}  \includegraphics[width=0.45\textwidth]{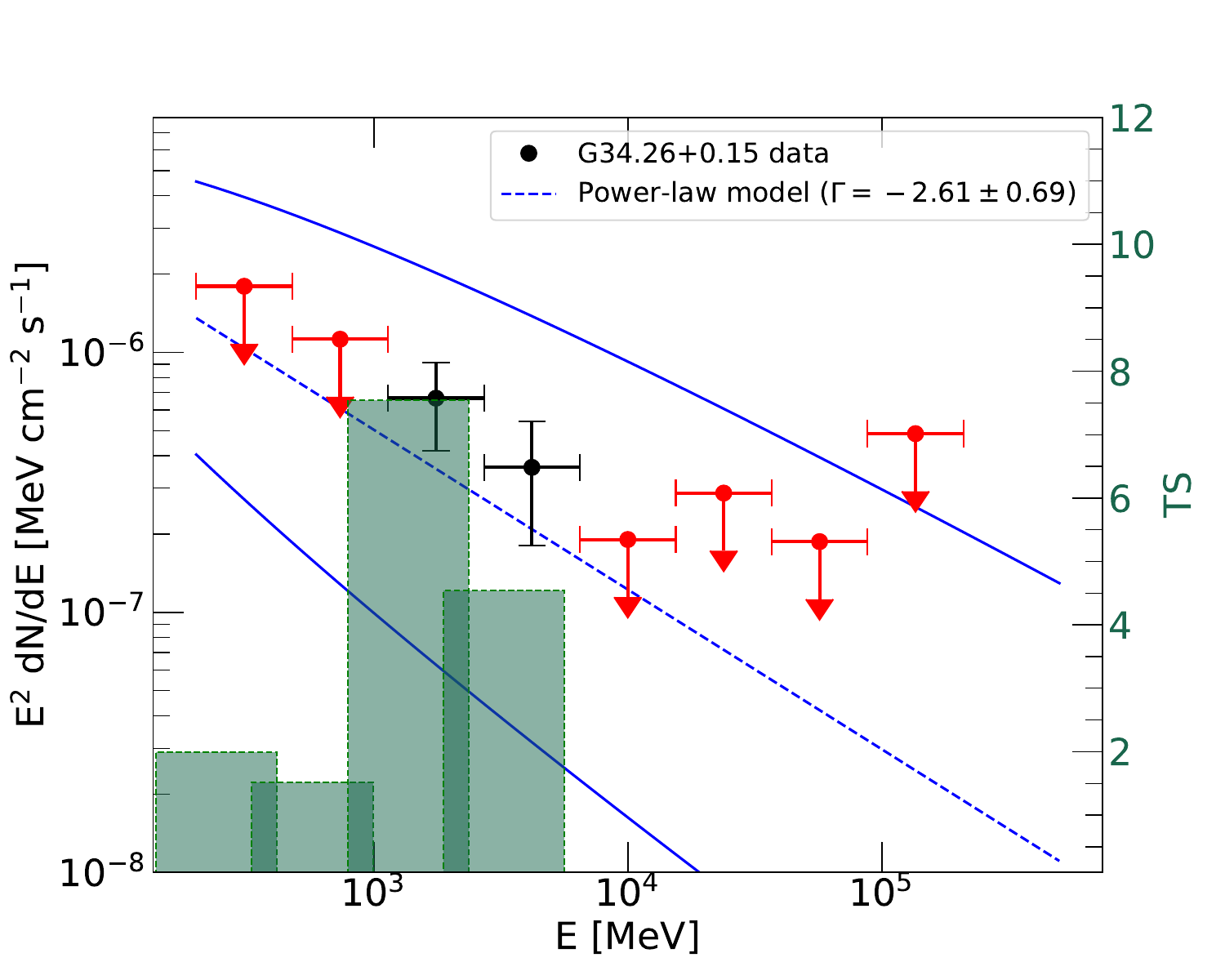} \\
   \includegraphics[width=0.45\textwidth]{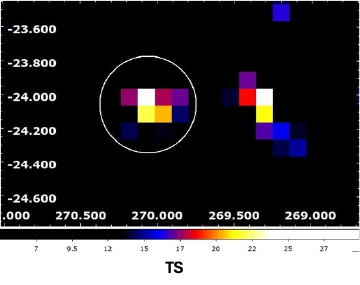}
   \includegraphics[width=0.45\textwidth]{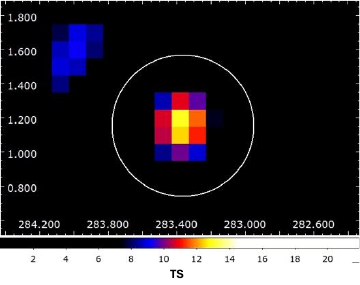}
    \caption{Top: the SEDs of G$5.89-0.39$ (left) and G$34.26+0.15$ (right) and their PL model fits. The red data points are the $2\sigma$ upper-limits, and the green histograms are the TS values of each flux bin. Bottom: the TS maps of G$5.89-0.39$ (left) and G$34.26+0.15$ (right) for the $2-500$ GeV energy range.}
    \label{fig:TS_sed_EDOs}
\end{figure*}

{For G$5.89-0.39$ modeled as a PL, we get the best-fit spectral parameters of $N_0 = (3.85 \pm 0.94)\times 10^{-15}$, $\Gamma = -2.12 \pm 0.13$, and $E_{\rm p} \simeq 14 \rm ~GeV$ (fixed parameter). Integrating this spectrum from $0.1-500$ GeV, we find the luminosity $L_\gamma \simeq (6.25\pm1.53)\times10^{34}$ erg s$^{-1}$. Finally, G$34.26+0.15$ PL model has best fit parameters of $N_0 = (5.04 \pm 3.11)\times 10^{-13}$, $\Gamma =  -2.61 \pm 0.45$, and $E_{\rm p} \simeq 10 ~\rm GeV$ (fixed parameter), leading to a luminosity $L_\gamma \simeq (1.05\pm0.65)\times10^{35}$ erg s$^{-1}$ over the $0.1-500$ GeV energy range. The TS maps ($2-500$ GeV) and the SEDs of these two sources are shown in Figure~\ref{fig:TS_sed_EDOs}.}

\section{Discussion}\label{sec:discussion}

In this section, we examine the potential association between the observed $\gamma$-ray emission and CR acceleration in EDOs. Section~\ref{sec:population} evaluates the collective contribution of EDOs to the Galactic cosmic-ray budget based on our population analysis. In Section~\ref{sec:DR21_gamma-ray}, we investigate the origin of the $\gamma$-ray emission in the DR21 region and assess possible emission scenarios.

\begin{deluxetable*}{clcc}
\tablenum{3}
\tablecaption{List of all the EDOs, their $\gamma$-ray luminosity and corresponding CR acceleration efficiency.
}
\tablewidth{240pt}
\tablehead{
\colhead{Source Name} &   
\colhead{$L_{\rm Outflow}$} &
\colhead{$\gamma$-ray luminosity\tablenotemark{$\alpha$}} &
\colhead{$\eta_{\rm CR}$\tablenotemark{$\alpha$}}}
\startdata
G$34.26+0.15$ & $1.67 \times 10^{36}$ erg s$^{-1}$ & $1.05\times10^{35}$ erg s$^{-1}$ & $18\%$ \\
DR21 & $3.17 \times 10^{36}$ erg s$^{-1}$  & $1.71\times10^{35}$ erg s$^{-1}$ & $15\%$ \\
Sh $106-$IR &  $9.07 \times 10^{35}$ erg s$^{-1}$ & $ < 4.41\times10^{32}$  erg s$^{-1}$ & $0.15\%$ \\
IRAS $16076-5134$ & $(0.9-9.1) \times 10^{37}$ erg s$^{-1}$  & $< 5.89\times10^{33}$ erg s$^{-1}$ & $0.02\%-0.2\%$  \\
G$5.89-0.39$ & $(0.003-31.81) \times 10^{37}$ erg s$^{-1}$ & $6.25 \times10^{34}$ erg s$^{-1}$ & $0.05\%-12\%$\\
IRAS $12326-6245$ & $4.54 \times 10^{37}$ erg s$^{-1}$ & $< 5.78 \times10^{35}$ erg s$^{-1}$ & $5\%$  \\
Orion BN/KL &  $6.34 \times 10^{36}$ erg s$^{-1}$ & $< 1.72 \times10^{34}$ erg s$^{-1}$ & $1\%$  \\
\enddata
\tablenotetext{a}{Calculated in this work. $\eta_{\rm CR}$ represents the upper limit on CR efficiency.}
\label{tab:results}
\end{deluxetable*}

\subsection{Population of EDOs}\label{sec:population}

{In Table~\ref{tab:table_sample}, we can see that most of the EDOs have a high total kinetic energy ($\gtrsim 10^{47}$~erg) and a high number density ($\gtrsim10^{4}$~cm$^{-3}$).} These conditions make it plausible that the observed $\gamma$-ray emission is produced by CR protons, accelerated by a shock produced in an explosive event. The CR protons then collide with dense gas to produce pions, and the neutral pions decay into $\gamma$-rays. 

Under these assumptions, we can evaluate the acceleration efficiency $\eta_{\rm CR}$, which tells us the fraction of outflow kinetic energy that goes into CR acceleration.  The efficiency can be approximated as
\begin{equation}
    \eta_{\rm CR} \simeq \frac{3L_{\gamma}}{L_{\rm outflow}},
\end{equation}
where $L_{\gamma}$ is the observed $\gamma$-ray luminosity (reported in Table~\ref{tab:results}), $L_{\rm outflow}$ is the total kinetic power of the EDO, and the factor of 3 is because only one-third of the CR created pions are neutral and decay into $\gamma$-ray photons. The expression assumes the calorimetric limit, where all CR protons collide with dense gas to produce pions, giving us an upper limit on the acceleration efficiency. To calculate $L_{\rm outflow}$, we adopt $L_{\rm outflow} = \rm KE/t_{\rm dyn}$, where $\rm KE$ is the kinetic energy and $t_{\rm dyn}$ is the dynamical age of the EDO. The latter is an upper limit since it is calculated based on a constant observed maximum radial velocity, leading to an upper limit on $\eta_{\rm CR}$ summarized in Table~\ref{tab:results}. The  $\eta_{\rm CR}$ upper limits vary between $0.01\%-18\%$ for the EDOs and are in accordance with the values predicted by theoretical estimates \citep[$\approx5\%$;][]{Araudo_2021}; \citep[$\approx10^{-4}\%-10^{-3}\%$;][]{Padovani2016}. 

Notably, DR21, G$34.26+0.15$ and G5.89$-$0.39, which are among the more evolved (older ages and high ambient densities) and energetically prominent outflows in the sample, exhibit higher $\gamma$-ray luminosities than the other sources. The observed $\gamma$-ray emission in these older systems may indicate efficient particle confinement or sustained interactions with dense surrounding material. This suggests that environmental conditions, such as the ambient density and/or diffusion coefficient associated with CRs, may play a critical role in shaping $\gamma$-ray detectability. For example, in \citet{Pandey2024}, we showed that in young massive star-forming regions, the observed $\gamma$-ray emission from the hadronic scenario depends on the balance between CR acceleration efficiency ($\eta_{\rm CR}$), diffusion, and ambient gas density. Faster CR escape requires a higher $\eta_{\rm CR}$ to sustain the observed emission. In contrast, higher gas densities increase interaction rates between CRs and dense gas, thereby reducing the impact of escape losses. These results suggest that hadronic models can explain the observed $\gamma$-ray luminosities in dense environments, provided CRs interact efficiently with surrounding material. We reach a similar conclusion for EDOs, but accurately constraining the scenario is challenging due to the lack of precise measurements of the gas number density in the EDOs reported in the literature.

The true frequency and origin of EDOs in the Milky Way remain poorly constrained. \citet{Guzm_IRAS5134} estimated a rate of one such event every 110 years. However, this rate should be considered a lower limit due to the limited number of detections and reliance on targeted observations. A complete and unbiased survey of massive star-forming regions, particularly those hosting known protostellar outflows, conducted with high-resolution and high-sensitivity millimeter facilities such as ALMA, would likely uncover a substantially larger population of EDOs and provide a more accurate estimate of their occurrence. 

The inferred EDO occurrence rate is comparable to that of core-collapse supernovae in the Galaxy, estimated at approximately one event every 50 years \citep{Tammann1994}. Although individual EDOs release roughly $10^{49}$ erg, about 100 times less than the $10^{51}$~erg typically emitted by a single supernova \citep{Hamuy2003}, the cumulative energy input from EDOs may still be non-negligible. For the same $\eta_{\rm CR} \simeq 10\%$ and if the current rate holds, EDOs could contribute at least $1\%$ of the total CR production by supernovae. This value could be potentially more, considering that many EDOs likely remain undetected. Moreover, \citet{Krumholtz2023} highlights the role of protostellar outflows as important local sources of CRs within star-forming regions. Their analysis indicates that while protostellar jets and accretion shocks are globally subdominant by contributing an order of magnitude less to the $\gamma$-ray emission than SNe, they may still be significant on local scales.

\begin{figure*}[t]
    \begin{center}
\includegraphics[width=0.82\textwidth]{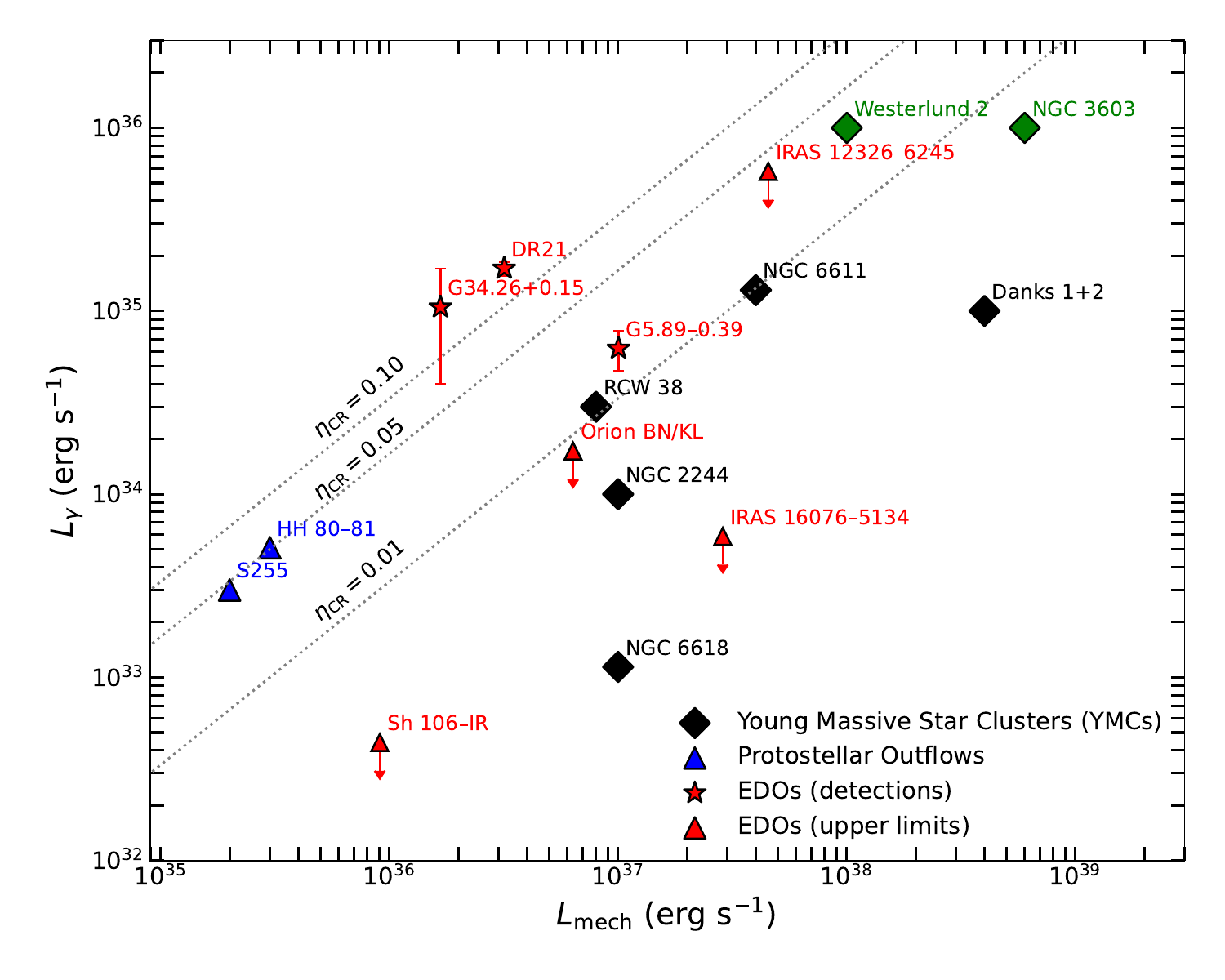}
       \caption{Gamma-ray luminosity ($L_\gamma$) versus mechanical power ($L_{\rm mech}$) for EDOs and young star-forming regions. Red stars indicate the three EDOs with confirmed $\gamma$-ray emission analyzed in this work. For comparison, black diamonds denote young massive star clusters (ages $\leq$3 Myr) and blue triangles represent protostellar outflows with confirmed Fermi-LAT $\gamma$-ray emission. The diagonal dotted lines mark constant CR acceleration efficiencies, $\eta_{\rm CR} = 3 L_\gamma$/$L_{\rm mech}$ = 0.01, 0.05, 0.10, illustrating approximate calorimetric limits for CR acceleration in dense star-forming systems. This figure places the EDOs in the broader context of young, massive star-forming regions and highlights that their $\gamma$-ray luminosities are comparable to or exceed those of many star clusters. Westerlund 2 and NGC 3603 are shown in green color because their $\gamma$-ray luminosity is calculated above 1 GeV, whereas for the rest of the sources, the values are reported between 0.2$-$500 GeV. Values taken from - S255: \citet{emma2023}, HH 80-81 \citet{Yan2022, Emma2025}, RCW38 - \citet{Pandey2024}, NGC2244 - \citet{Liu2023}, NGC6618 - \citet{Liu22, Rosen_2014}, NGC6611 - \citet{Peron25_NGC}, Danks 1 and 2 - \citet{Liu24_danks}, Westerlund2 - \citet{Westerlund2}, and NGC3603 - \citet{NGC36031, NGC36032, Rosen_2014}. } 
       \label{fig:all_sfr}
    \end{center}
\end{figure*}

{To place the EDOs in the broader context of young massive star-forming environments as $\gamma$-ray emitters, in Figure~\ref{fig:all_sfr} we compare the $\gamma$-ray luminosity $L_\gamma$ to the inferred mechanical power $L_{\rm mech}$ for several classes of sources younger than 3 Myr. The EDOs analyzed in this work are denoted by red stars (red triangles for upperlimits) and their approximate uncertainties in $\gamma$-ray luminosity are shown. For comparison, we include well-studied young massive clusters (black diamonds) and the two protostellar outflows detected in Fermi-LAT data (blue triangles), each of which occupies a distinct region in the $L_\gamma-L_{\rm mech}$ space. The dotted diagonal lines represent constant acceleration efficiencies, $\eta_{\rm CR}=$ 0.01, 0.05, and 0.10, which approximate the calorimetric limits expected when CRs are efficiently confined and lose most of their energy to hadronic interactions. This comparison emphasizes that EDOs occupy a physically distinct regime between protostellar jets and massive clusters, reinforcing the idea that they may represent an important source of high-energy emission in the early stages of massive star formation.}

\subsection{Origin of Gamma-ray emission from DR21}\label{sec:DR21_gamma-ray}

{In this section, we examine physical mechanisms responsible for the observed $\gamma$-ray emission associated with DR21 as the most $\gamma$-ray luminous EDO. We compare our results with previous observational studies and theoretical models of the region, and we assess the possibility of different emission scenarios given our Fermi-LAT results.}

\subsubsection{Previous Fermi-LAT Studies on DR21 and the Cygnus-X region}

DR21 is a prominent site of massive star formation within the Cygnus-X giant molecular cloud complex \citep{Schneider2006}. Cygnus X is a $\sim7^{\circ} \times 7^{\circ}$ star-forming complex in the Cygnus constellation, centered near Gamma Cygni, the star in the Northern Cross \citep{Reipurth2008}. Cygnus X is the most active star-forming region within 2 kpc of the Sun, hosting $\sim800$ H\,\textsc{ii} regions, Wolf-Rayet and O stars, several OB associations, over 40 massive protostars, and a molecular cloud complex of $\sim3\times10^{6}\,M_{\odot}$ \citep{Wendker1984, Wendker1991, Schneider2006, Schneider2007}. Spitzer IRAC and MIPS photometry further revealed a rich young stellar population, including 670 Class I, 7249 Class II, 112 transition disks, and 200 embedded protostars, with $58-67\%$ of young stellar objects (YSOs) clustered in groups $\geq$10 members, particularly south-west of the DR21 region \citep{Beerer2010}.

The Cygnus-X region has been extensively studied previously using Fermi-LAT observations \citep{Ackerman_sci_cyg, 2012Ack_cyg, Aharoniannature}. \cite{Ackerman_sci_cyg} reported the discovery of a 50-parsec-wide ``cocoon" of freshly accelerated CRs within the Cygnus-X region in the 1$-$100 GeV range, particularly in the vicinity of the massive Cygnus OB2 stellar cluster using the Fermi LAT. The region is filled with high-energy particles, possibly accelerated by the combined effects of stellar winds and supernova activity from young massive stars. The most recent, extensive study on the Cygnus-X region was conducted by \cite{Cyg2}. Using over 13 years of Fermi-LAT data, they performed a morphological analysis of $\gamma$-ray emission in Cygnus-X and found that the ``Cygnus cocoon” comprises three spatially distinct, extended components, and the cocoon's morphology is best described by overlapping structures. In our work, we adopt the standard Fermi-LAT analysis framework with the 4FGL-DR4 catalog. We note that the results presented here are not sensitive to the specific large-scale cocoon morphology adopted in previous studies. 

\subsubsection{The Explosive Outflow in DR21}\label{sec: sample_DR21}

DR21 is located at a distance of $1.50^{+0.08}_{-0.07}$~kpc based on trigonometric parallax measurements \citep{Rygl2012}. DR21 hosts several compact H{\sc ii} regions and deeply embedded massive protostars and is known for driving one of the most massive and luminous molecular outflows in the Milky Way \citep{Beerer2010, Zapata2013}. The core mass of DR21 is estimated to exceed $2\times10^4$\,$M_\Sun$ \citep{Richardson1989, White2010}.  The outflow associated with DR21 is highly energetic with a luminosity in the 2 $\mu$m band alone exceeding 1800\,$L_\Sun$ \citep{Garden1991, Garden1996}. Recent high-resolution ALMA observations by \cite{Guzman2024ALMA} have confirmed the presence of an EDO within DR21, traced by more than a dozen high-velocity CO(2–1) streamers that converge on a common center. 

Figure~\ref{fig:DR21_multiwavelength} shows the DR21 ridge, which is an elongated, filamentary structure that includes two major cores: DR21(OH) in the north and DR21 in the south. DR21(OH) is an active, high-mass star-forming region characterized by strong maser emission and compact millimeter sources \citep{Cao2022}. Just north of the ridge lies W75N, another massive star-forming region within the Cygnus-X complex. Although DR21 and W75N appear close in projection, they are distinct in both spatial position and radial velocity \citep{rygl2010, Rygl2012}, indicating they are physically separate systems.

\begin{figure}[t]
\centering
\includegraphics[width=\columnwidth]{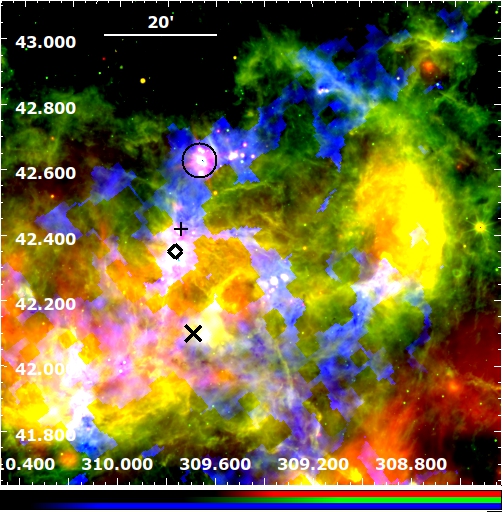}
\caption{Multiwavelength image of the DR21 region: 22$\mu$m allWISE (red), 12$\mu$m allWISE (green), and CN emission (blue). The black circle marks W75N (a massive star-forming region), the black cross highlights the DR21(OH) maser, the diamond marks the DR21 EDO, and the X represents the 4FGL $\gamma$-ray counterpart of the EDO. North is up, and East is to the left.
}
\label{fig:DR21_multiwavelength}
\end{figure}

\subsubsection{DR21 as a gamma-ray source}

{The energetic requirements of the observed $\gamma$-ray emission support an origin from the explosive outflow. In Fig.~\ref{fig:tsmap_3color} (a) and in the associated analysis in Section~\ref{sec:likelihood_DR21}, we show that the $\gamma$-ray source is spatially coincident with the EDO DR21. Figure~\ref{fig:tsmap_3color}(b) also demonstrates that the region is embedded in very dense gas with number densities of order $\sim 10^{5}$~cm$^{-3}$, as traced by CN emission. These environmental conditions are ideal for efficient hadronic interactions following local CR acceleration. Taken together, the kinetic energy budget, the typical CR acceleration efficiency observed in young star-forming regions, and the spatial correlation, the evidence favors local particle acceleration driven by the DR21 explosive outflow as the dominant origin of the observed $\gamma$-ray emission.}

{The DR21 region hosts bright mid-IR sources, but deep near-IR cluster studies do not reveal a substantial number of main-sequence O-type stars. In the wide-field DR21/W75 survey of \citet{Davis2007}, only one object in the field was considered a possible O-type candidate, and this identification remains uncertain. Thus, the confirmed massive-star content in DR21 is sparse as it is still in the star formation process, removing the possibility that the observed emission could be a product of CR acceleration by stellar winds.}

{As another possible source, the halos of superluminous or massive stars have been explored both theoretically and observationally as gamma-ray emitters \citep[e.g.][]{Orlando2007, Bednarek_2021, de_Menezes_2021}. However, deep Fermi-LAT observations show that such systems are generally faint or undetected at GeV energies. Given the absence of confirmed superluminous stars or a rich OB population in DR21, yet in its early star formation phase, this scenario is therefore unlikely in the present case.}

{The mid-IR survey of DR21 reveals more than 50 individual H$_2$ knots and bow shocks distributed along the DR21 and DR21(OH) filaments. Bow shocks of this type are produced by high-velocity protostellar jets and outflows interacting with dense molecular gas and are natural sites of diffusive shock acceleration (DSA) \citep{2015Padovani, Padovani2016}. More recently, the study by \citet{Karska2025sofia} shows that the DR21 EDO also drives multiple internal shocks along its $\sim$1~pc-scale cavity. Their shock diagnostics, such as broadened CO line wings and H$_2$ excitation, indicate shock speeds of  $\approx$~20~km~s$^{-1}$, fully within the regime capable of accelerating particles to GeV energies in dense environments. Using Hillas Criteria and Equation~7 from \citet{owen2023}, the maximum energy attained by CRs during the acceleration process can be estimated from the shock velocity $v_{\rm sh}$, the size of the acceleration region  $r$, and its mean magnetic field strength, $\langle |B| \rangle$:}

\begin{equation}
E_{\max} \sim 1 \left( \frac{r}{\mathrm{pc}} \right)
              \left( \frac{v_{\rm sh}}{1{,}000~\mathrm{km\,s^{-1}}} \right)
              \left( \frac{\langle |B| \rangle}{\mu\mathrm{G}} \right)
              \ \mathrm{TeV} .
\end{equation}

{For the EDO DR21, $v_{\rm sh}=20$~km~s$^{-1}$ \citep{Karska2025sofia}, $r\approx1$~pc \citep[DR21 filament length;][]{Schneider2006}, and $\langle |B| \rangle= 0.6$~mG \citep{Ching_2022}. Using these values, we get $E_{\rm max} \approx 12$~TeV for accelerated CRs.  Assuming that a gamma-ray photon carries about 10$\%$ of the proton energy, $E_{\rm max} \approx 1$~TeV, which is in accordance with our observational constraints of DR21 (see spectrum in Figure~\ref{fig:sed_dr21}).}

{An alternative to local particle acceleration in the DR21 explosive outflow is that the $\gamma$-ray emission arises from interactions between dense gas and pre-existing CRs associated with the Cygnus cocoon. In this context, \citet{menchiari2024} modeled the expected $\gamma$-ray emission from DR21 assuming illumination by particles accelerated at the Cyg~OB2 wind-termination shock, with DR21 acting as a passive target. While we cannot exclude the possibility that CRs from the Cygnus cocoon contribute at some level, a comparison between the predicted spectra in \citet{menchiari2024} and our observations reveals significant differences, most notably the presence of a spectral cutoff at $\sim10$~GeV in DR21 that is not reproduced by the model. Moreover, the viability of efficient particle acceleration at a cluster-scale wind-termination shock in Cyg~OB2 remains uncertain. Recent hydrodynamic simulations by \citet{vieu2024} suggest that the stellar association is too spatially unbound to sustain a coherent collective wind, potentially inhibiting the formation of a global termination shock.
Taken together, these considerations indicate that while some of the CRs in the region are accelerated by the Cygnus cocoon, it is unlikely to account for the observed $\gamma$-ray emission.}

\section{Conclusions} \label{Sec:Conclusions}

In this work, we have carried out the first systematic population study of EDOs in the context of GeV $\gamma$-ray emission. Using 16 years of Fermi-LAT data, we analyzed seven known EDOs, deriving either $\gamma$-ray luminosities or upper limits. Among the full sample, we detect $\gamma$-ray emission spatially consistent with three EDOs, namely, DR21, G34.26$+$0.15, and G5.89$-$0.39. While the remaining sources, Sh $106-$IR, IRAS $16076-5134$, IRAS $12326-6245$, Orion BN/KL, yield non-detections and upper limits. These findings highlight significant diversity in the high-energy properties of EDOs, ranging from objects with $\gamma$-ray luminosities of $\sim 10^{34}$–$10^{35}~$erg~s$^{-1}$ to systems whose luminosities are constrained to below $10^{33}$~erg~s$^{-1}$.

From these measurements, we estimated the corresponding CR acceleration efficiencies, $\eta_{\rm CR}$, assuming hadronic interactions within the dense, shocked molecular environment. The efficiencies inferred for a calorimetric limit for the detected sources span $\sim$0.01–18$\%$, broadly consistent with expectations for strong shocks interacting with dense gas. The wide range of efficiencies may reflect intrinsic variations in shock velocities, densities, magnetic-field strengths, and environmental confinement conditions across the sample. Assuming a CR acceleration efficiency of $\eta_{\rm CR}\sim10\%$ and using the currently estimated event rate, which is comparable to Galactic supernovae, EDOs could account for at least $\sim1\%$ of the Galactic CR budget normally attributed to supernovae. This fraction may be higher if a substantial number of EDOs remain undiscovered.

DR21 stands out as the brightest $\gamma$-ray emitter in the population, with  $L_\gamma \simeq (1.71\pm0.15)\times10^{35}$ erg s$^{-1}$ and an inferred $\eta_{\rm CR}$ of 15$\%$, under calorimetric assumption. Motivated by its luminosity and well-characterized molecular environment, we examined DR21 in greater detail. We discussed several plausible $\gamma$-ray production scenarios, including hadronic interactions from the EDOs and emission from the nearby superbubble Cygnus OB2. Among these, the energetics, morphology, and shock conditions strongly favor a scenario in which the explosive outflow itself accelerates CRs, which then interact with the dense molecular gas to produce the observed GeV $\gamma$-ray emission.

Overall, our study demonstrates that EDOs can be efficient and potentially significant producers of high-energy emission in young star-forming regions. The detection of $\gamma$-ray emission from three systems, combined with physically meaningful upper limits for the remainder, establishes EDOs as a possible class of energetic environments worthy of further study. Future high-sensitivity and wide-field surveys directed towards young massive star-forming regions will be essential to uncover additional EDOs and improve our understanding of their role in the Galaxy’s energy budget and their potential contribution to CR acceleration.

\begin{acknowledgements}
PP is grateful to the participants of the TOSCA workshop held at Siena (IT) for their insightful discussions. PP is thankful to Chris Kochanek, Ellis Owen, and David Smith for their feedback. TAT is supported in part by NASA grant 80NSSC23K1480. PP, SCL, and LAL acknowledge support through the Heising-Simons Foundation grant 2022-3533. TL is supported by the Swedish Research Council under contract 2022-04283, the Swedish National Space Agency under contract 117/19, and the Wenner-Gren Foundation under grant SSh2024-0037. Parts of this research were supported by the Australian Research Council Discovery Early Career Researcher Award (DECRA) through project number DE230101069.2407522. SSRO acknowledges support from NSF AST-2107340, a Peter O'Donnell Distinguished Researcher Fellowship, and a Donald Harrington Fellowship.
\end{acknowledgements}

\software{\texttt{FermiPy} Python package \citep[v1.4.0;][]{Wood_fermipy}, and ScienceTools version 2.2.0.}
\newpage
\bibliography{DR21}{}
\bibliographystyle{aasjournal}

\end{document}